# Survey of Operating Systems for the IoT Environment

**Tuhin Borgohain***
Department of Instrumentation Engineering, Assam Engineering College, Guwahati, India
Email: borgohain.tuhin@gmail.com
**Uday Kumar**
Tech Mahindra Limited, Chennai, India
Email: udaykumar@techmahindra.com
**Sugata Sanyal**
Corporate Technology Office, Tata Consultancy Services, Mumbai, India
Email: sugata.sanyal@tcs.com
*Corresponding Author*

------------------------------------------------------------------ABSTRACT-------------------------------------------------------------
This paper is a comprehensive survey of the various operating systems available for the Internet of Things environment. At first the paper introduces the various aspects of the operating systems designed for the IoT environment where resource constraint poses a huge problem for the operation of the general OS designed for the various computing devices. The latter part of the paper describes the various OS available for the resource constraint IoT environment along with the various platforms each OS supports, the software development kits available for the development of applications in the respective OS'es along with the various protocols implemented in these OS'es for the purpose of communication and networking.

## I. INTRODUCTION

The whole Internet of Things environment is based on the application of microprocessors and wireless sensors. The resource constraint environment of these microprocessors and sensors makes the use of regular OS'es meaningless due to their high resource and computing power requirement. Thus, in such a situation, the development of OS'es meeting the resource constraint demand of the IoT environment becomes necessary.

## II. OVERVIEW

In section III the paper introduces the various aspects of an OS designed for the IoT environment. In section IV, the various OS'es available for running in the IoT environment along with a list of the supporting platforms, SDKs and the various networking and communication protocols implemented are surveyed. The paper is concluded in section V.

## III. INTRODUCTION TO OS FOR THE IOT ENVIRONMENT

The whole integration of the various IoT devices to the various objects is made possible through the interaction of software at a dynamic level along with the use of wireless sensor network and RFID technologies using the internet infrastructure ([1], [6]). This software interaction is made possible through the operating system running behind the scene within each IoT device without which an IoT device would be nothing more than a non-functioning device. The flexible features of the various operating systems of an IoT device has facilitated some interesting integration of electronic products and technologies to the daily processes of an individual thus making the processes a whole lot easier to use and access. Some out of the multitudes of IoT technology integration and innovations are smart light bulb ([28]), implementation of real time passenger information system ([22]), smart tags/NFC tags ([17]) etc.

The OS'es developed for the IoT environment require very few kilobytes of RAM as well as operate with low power consumption. Moreover they are specifically designed and optimized for a particular set of microprocessor-based platforms beyond which such OS'es becomes irrelevant in its application ([38]). These OS'es do not compromise in terms of features relating to communication, networking, security etc. as compared to the regular OS'es like Windows OS, Mac OS etc. but comes built-in with a number of pre-installed, pre-integrated applications, drivers and other network protocols. Moreover these OS'es employ a number of unique security measures for enhancing the IoT infrastructure as a whole and to avoid the compromise of the stability and usability of the OS.

Though the security issues of the OS'es for the IoT environment are quite different in comparison to the security issues of a regular operating system, yet it still retains the standard security protocols for protecting itself against unwanted attacks. Now the IoT environment is made in such a way so as to carry out information exchange between the various electronic devices over the internet in the most efficient way possible using the lowest amount of resources. As such the whole IoT environment along with its OS becomes prone to malicious attack from the third party intruders. So the successful implementation of various encryption and data hiding techniques ([4], [5], [12], [15], [39]), intrusion detection systems ([16], [33]) etc. in the IoT infrastructure takes a paramount importance. [45] takes care of sleep deprivation attack blockage on IoT elements, to preserve their already fragile power resources.

## IV. OS'ES

i. mbed: Developed by ARM in collaboration with its technological partners, mbed OS is developed for 32-bit ARM Cortex-M microcontrollers ([29]). The whole OS is written using C and C++ language. This open source OS is licensed under Apache License 2.0.

The software development kit (SDK) for mbed OS provides the software framework for the developers to develop various microcontroller firmwares to be run on IoT devices. These SDK is comprised of core libraries which consist of the following components given in Table 1:

| Networking | Test scripts | Microcontroller peripheral drivers | RTOS and runtime environment | Build Tools | Debug Scripts |
|---|---|---|---|---|---|

Table 1: Core libraries in mbed OS

The applications for mbed OS can only be developed online using its native online code editor cum compiler known as mbed online integrated development environments (IDEs). While writing of code can only be done through a web browser, its compilation is done by the ARMCC C/C++ compiler in the cloud.

In the connectivity front, the mbed OS support the following connectivity technologies given in Table 2:

| Bluetooth Low Energy | Wi-fi | Zigbee IP | Zigbee LAN |
|---|---|---|---|
| Cellular | | Ethernet | 6LoWPAN |

Table 2: Connectivity technologies in mbed OS

mbed OS integrates end-to-end IP security (IPv4 and IPv6) through TLS and DTLS in its comm. channels for increased security of the whole OS environment. Moreover for management of various devices in its environment, mbed OS uses OMA Lightweight M2M protocol.

ii. RIOT: Developed by INRIA, HAW Hamburg and FU Berlin initially, RIOT OS is compatible with ARM Cortex-M3, ARM Cortex-M4, ARM7, AVR Atmega and TI MSP430 devices ([8], [24], [31]). Developed using C and C++, this open source OS is licensed under LGPL v2.1.

The SDKs available for development of applications in RIOT OS are gcc, valgrind and gdb. Moreover the SDK framework supports application programming in C and C++.

RIOT OS supports all the major communication and networking protocols which are tabulated in Table 3:

| IPv6 | 6LoWPAN | RPL | CoAP |
|---|---|---|---|
| UDP | TCP | CBOR | CCN-lite |
| OpenWSN | | UBJSON | |

Table 3: Networking protocols in RIOT OS

iii. Contiki: Created by Adam Dunkels and further developed by people from various organisations and institutions like Atmel, Cisco, ENEA, SAP, Sensinode, Oxford University etc. ([3], [19], [35]), the Contiki OS is aimed to be used in various microcontroller devices which are tabulated in Table 4:

| Atmel ARM | Atmel AVR | STM32w | TI MSP430 |
|---|---|---|---|
| TI CC2430 | TI CC2538 | TI CC2630 | TI CC2650 |
| LPC2103 | Freescale MC13224 | Microchip dsPIC | Microchip PIC32 |

Table 4: Microcontroller devices running on Contiki OS

This open source OS is licensed under BSD License.

The programming model of the Contiki OS is based on protothreads for efficient operation in resource-constrained environment.

The Contiki OS features Cooja, a network simulator which simulates Contiki nodes ([18]). These Contiki nodes are of three types:
 a. Emulated nodes
 b. Cooja nodes
 c. Java nodes

The various networking protocols supported by Contiki OS are given in Table 5:

| CoAP | 6LoWPAN | RPL |
|---|---|---|

Table 5: Networking protocols in Contiki OS

The Contiki environment is generally made secure through the implementation of ContikiSec ([21]) and through the implementation of TLS/DTLS ([9]).

iv. TinyOS: Developed by TinyOS Alliance, this open source OS is mainly developed for wireless sensor networks ([2], [7]). It is written in nesC and is licensed under BSD License.

The SDK for application development for TinyOS is comprised of the following three IDEs:
 a. TinyDT
 b. TinyOS Eclipse Plugin "YETI 2"
 c. TinyOS Eclipse Editor Plugin

The various communications and network protocols implemented in the TinyOS are given in Table 5:

| Broadcast based Routing | Probabilistic Routing | Multi-Path Routing |
|---|---|---|
| Geographical Routing | Reliability based Routing | TDMA based Routing |
| Directed Diffusion | | |

Table 5: Communication and networking protocols in TinyOS

The whole architecture of the TinyOS has been made secure over the years with the implementation of TinySec ([44]) and various types of embedded security layers ([13], [14], [30], [40]).

v. Nano-RK: Developed at Carnegie Mellon University by Alexei Colin, Christopher Palmer and Artur Balanuta, Nano-RK is specifically targeted for running in microcontrollers (presently runs on MicaZ motes and FireFly Sensor Networking Platform) to be used in wireless sensor networks. Nano-RK OS is written in C language and is open source ([10], [43]).

The application development of Nano-RK OS is supported by the Eclipse IDE.

The communications within the OS is carried out with the help of the following protocols given in Table 6:

| RT-Link | PCF TDMA | b-mac | U-Connect | WiDom |
|---|---|---|---|---|

Table 6: Communication protocols implemented in Nano-RK

vi. FreeRTOS: Developed by Real Time Engineers Ltd., the FreeRTOS is developed for platforms listed in Table 7:

| ARM7 | ARM9 | ARM Cortex-M3 | ARM Cortex-M4 |
|---|---|---|---|
| ARM Cortex-A | RM4x | TMS570 | Cortex-R4 |
| Atmel AVR | AVR32 | HCS12 | Altera Nios II |
| MicroBlaze | Cortus APS1 | Cortus APS3 | Cortus APS3R |
| Cortus APS5 | Cortus FPF3 | Cortus FPS6 | Cortus FPS8 |
| Fujitsu MB91460 series | Fujitsu MB96340 series | Coldfire | V850 |
| 78K0R | Renesas H8/S | MSP430 | 8052 |
| X86 | RX | SuperH | PIC |
| Atmel SAM3 | Atmel SAM4 | Atmel SAM7 | Atmel SAM9 |

Table 7: Platforms supporting FreeRTOS

Written mostly in C with the addition of a few assembly functions, this open source OS is licensed under Modified GPL ([25], [26]).

The application development part for FreeRTOS is handled through multiple threads, software timers and semaphores along with a tick-less mode for low consumption of resources by the running of the various applications.

V. CONCLUSION

From the above survey it can be seen that all the OS'es for the IoT environment are well equipped with all the major networking and communication protocols, security features as well as optimized for efficient usage of computing power in a resource constraint environment. Yet the additional implementation of counter measures to online dictionary attacks ([11], [20], [32]) in the internet infrastructure used by the IoT environment with the additional emphasis on developing a more robust wireless sensor network ([27], [34], [37], [41]) will contribute to the protection of user's credentials during online transactions ([23], [36], [42]) logging inside one's personal account in the cloud and will make the whole IoT environment much secure and more reliable.

networks." In *Real-Time Systems Symposium, 2005. RTSS 2005. 26th IEEE International*, pp. 10-pp. IEEE, 2005.

[44] Chris Karlof, Naveen Sastry, and David Wagner. "TinySec: a link layer security architecture for wireless sensor networks." In *Proceedings of the 2nd international conference on Embedded networked sensor systems*, pp. 162-175. ACM, 2004.

[45] Tapalina Bhattasali, Rituparna Chaki, Sugata Sanyal; "Sleep Deprivation Attack in Wireless Sensor Network": International Journal of Computer Applications, Volume 40,Number 15, pp.19-25, February 2012, ISBN: 978-93-80866-55-8, Published by Foundation of Computer Science, New York, USA, DOI: 10.5120/5056-7374.

**Biographies and Photographs**

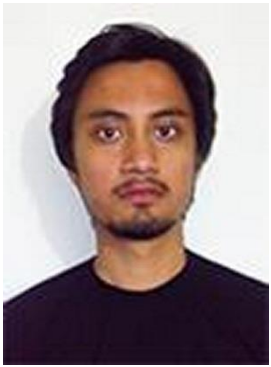

Tuhin Borgohain is a 3rd Year student of Assam Engineering College, Guwahati. He is presently pursuing his Bachelor of Engineering degree in the department of Instrumentation Engineering.

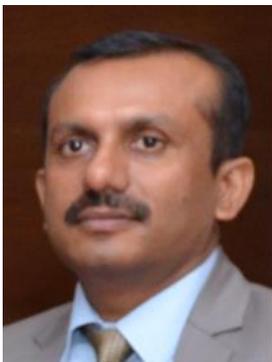

Uday Kumar is working as Delivery Manager at Tech Mahindra Ltd, India. He has 17 years of experience in engineering large complex software system for customers like Citibank, FIFA, Apple Smart objects and AT&T. He has developed products in BI, performance testing, compilers. And have successfully led projects in finance, content management and ecommerce domain. He has participated in many campus connect program and conducted workshop on software security, skills improvement for industrial strength programming, evangelizing tools and methodology for secure and high end programming.

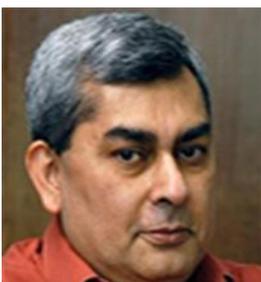

Sugata Sanyal is presently acting as a Research Advisor to the Corporate Technology Office, Tata Consultancy Services, India. He was with the Tata Institute of Fundamental Research till July, 2012. Prof. Sanyal is a: Distinguished Scientific Consultant to the International Research Group: Study of Intelligence of Biological and Artificial Complex System, Bucharest, Romania; Member, "Brain Trust," an advisory group to faculty members at the School of Computing and Informatics, University of Louisiana at Lafayette's Ray P. Authement College of Sciences, USA; an honorary professor in IIT Guwahati and Member, Senate, Indian Institute of Guwahati, India. Prof. Sanyal has published many research papers in international journals and in International Conferences worldwide: topics ranging from network security to intrusion detection system and more.